\PassOptionsToClass{balance=false}{acmart}
\documentclass[sigconf,nonacm]{acmart}
\setcopyright{none}
\renewcommand\footnotetextcopyrightpermission[1]{}
\acmDOI{}
\acmISBN{}
\acmConference[WWW '27]{Proceedings of the ACM Web Conference 2027}{May 10--14, 2027}{Dublin, Ireland}
\acmBooktitle{Proceedings of the ACM Web Conference 2027 (WWW '27), May 10--14, 2027, Dublin, Ireland}
\acmYear{2027}
\copyrightyear{2027}
\usepackage{booktabs,tabularx,amsmath,graphicx,microtype,multirow}
\newif\ifarxivversion
\arxivversiontrue
\newcommand{\OneRecCVCondWNPoint}{-0.000887}
\newcommand{\OneRecCVCondWNLo}{-0.002209}
\newcommand{\OneRecCVCondWNHi}{+0.001211}
\newcommand{\OneRecCVBootstrapDraws}{5,000}
\begin{document}
\title{Training with Missed Targets in Generative Recommendation: Separating Supervision from Probability Competition}
\author{Xuesi Wang}
\authornote{Xuesi Wang and Yangbin Shi contributed equally to this work.}
\authornote{Corresponding author.}
\email{wxsthu@icloud.com}
\affiliation{%
  \institution{Independent Researcher}
  \city{Shanghai}
  \country{China}}

\author{Yangbin Shi}
\authornotemark[1]
\email{11921182@zju.edu.cn}
\affiliation{%
  \institution{Zhejiang University}
  \city{Hangzhou}
  \country{China}}

\author{Xiaolin Zheng}
\email{xlzheng@zju.edu.cn}
\affiliation{%
  \institution{Zhejiang University}
  \city{Hangzhou}
  \country{China}}
\renewcommand{\shortauthors}{Wang, Shi, and Zheng}
\begin{abstract}
Generative recommenders return a limited candidate set and may omit observed targets before reranking. A training strategy appends these missed targets to reranker training lists, although inference still ranks only original candidates. This operation simultaneously changes retrieved-target weight, adds supervision over appended targets, and makes the two groups compete for probability. An append/no-append comparison therefore cannot explain changes in returned-item rankings. We construct three matched losses that hold retrieved-target weight fixed while introducing appended-target supervision and group competition separately. The intermediate loss trains within both groups but normalizes them separately, preventing training-only targets from competing with inference candidates. Experiments with a released OneRec model and locally trained Amazon generators show that this competition can harm returned-item ranking. In four prespecified Amazon Video Games comparisons, removing it improved full-target normalized discounted cumulative gain (FT-NDCG) by 7.8--22.2\%; 95\% intervals over users and three of four intervals over training runs excluded zero. A conservative development-set rule selected appended-target training for two of three generators in one held-out category and rejected it for all three in another, avoiding a 1.7\% loss. Candidate completion should therefore be evaluated for each generator rather than applied automatically.
\end{abstract}
\maketitle
\section{Introduction}
\label{sec:introduction}

Large Web recommenders separate candidate generation from ranking because a changing catalog cannot be scored exhaustively within a serving-time budget~\cite{covington2016youtube}. Generative recommenders retrieve items by producing \emph{semantic identifiers} (semantic IDs), short token sequences assigned to catalog items~\cite{rajput2023tiger,hua2023index}. They may replace a conventional cascade~\cite{wang2026unirec} or provide a quota-limited channel blended with other retrieval routes~\cite{yang2026complementary}; a reranker or verifier can refine that channel's slate~\cite{yang2025liger,zhang2026verifier}. Once the generator and its candidate route are held fixed, an omitted item cannot reach its reranker. This creates a Web recommendation mismatch: feedback spans a changing catalog, whereas each serving route exposes only a bounded slate.

One way to use labels missed by the generator is \emph{candidate completion}: keep every retrieved item, append the missed observed targets to the reranker's training list, and train on the enlarged list. The appended targets may come from another retrieval route or from an offline answer set. They are absent again at inference, when the reranker receives only the generator's original candidates. The practical question is whether learning against these training-only items improves or harms the ranking of items that can actually be served.

\begin{figure*}[t]
\centering
\includegraphics[width=\textwidth]{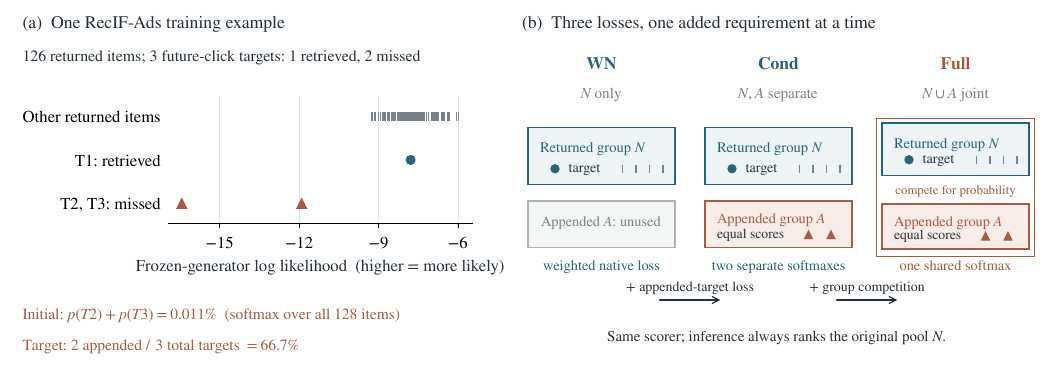}
\caption{Candidate completion and three training losses. (a) One RecIF-Ads answer set contains three future-click targets. T1 appears among 126 returned items, while T2 and T3 are appended only for training. Together, T2 and T3 initially receive 0.011\% of the softmax probability over all 128 training items; uniform supervision assigns them $2/(1+2)=66.7\%$ of target weight. (b) WN retains only weighted ranking within the returned group; Cond adds a separately normalized appended-target task; Full uses one normalization over both groups and thereby adds cross-group competition. Thus WN$\rightarrow$Cond and Cond$\rightarrow$Full each add one training requirement.}
\Description{Panel a plots frozen-generator scores for 125 returned nontargets, one returned target, and two missed targets, then shows how the initial appended-group probability and its uniform target weight are calculated. Panel b compares WN, Cond, and Full by showing which candidates enter training, which items are ranked within each group, and whether the two groups compete for probability.}
\label{fig:training_pool}
\end{figure*}

Figure~\ref{fig:training_pool}(a) shows the mismatch in one example from the public RecIF-Bench Ads recommendation dataset (RecIF-Ads)~\cite{zhou2026openonerec}. Its answer set contains three future-click targets. The generator returns 126 items containing the first target and misses the other two, which are appended only for training. Although the appended pair initially holds 0.011\% of the model probability, the uniform completed-list loss asks it to receive 66.7\%. This creates strong competition with returned items that does not exist at inference.

Figure~\ref{fig:training_pool}(b) compares three losses. Weighted-native (WN) ranks retrieved targets against the other returned items, then weights this ranking loss by the fraction of all observed targets that the generator retrieved. Conditional training (Cond) adds a separately normalized loss over appended targets, without competition between the groups. Full append training (Full) puts both groups in one softmax, adding competition to the same two within-group tasks. We derive Cond by allowing all appended scores to move together and choosing the shared shift that minimizes Full. This removes only the requirement that the two groups compete for total probability. Comparing Cond with WN tests the appended-target loss; comparing Cond with Full tests removal of group competition. All three use the same training users and inference pool. With one appended target, its within-group probability is 1 and its loss is zero, so Cond reduces to WN; 63.5--100\% of our training groups contain at least two appended targets.

Earlier work shows that candidate distributions affect ranking cascades~\cite{qin2022rankflow} and that adding gold references can harm machine-translation rerankers~\cite{bhattacharyya2021energy}. Neither result explains what happens when observed targets are appended only to a recommender's training lists. This operation lowers the target weight assigned to retrieved items, adds supervision over the appended targets, and makes the two groups compete for probability. We match the retrieved-target weight across all three losses, then use the WN$\rightarrow$Cond$\rightarrow$Full path to introduce appended-target supervision and group competition separately. Table~\ref{tab:closest_work} distinguishes this controlled comparison from nearby methods.

We train WN, Cond, and Full with the same candidates, representations, scorer architecture, optimizer, and inference pool. Comparing their returned rankings attributes a change to appended-target supervision or to group competition. Evaluation includes all eligible users, including those for whom the generator retrieved no target. We then address the practical choice directly: development users determine whether to keep returned-only training or adopt appended-target training, and different users from held-out categories evaluate that choice.

Our contributions are threefold. First, we show that candidate completion changes retrieved-target weight, adds supervision over missed targets, and creates probability competition with returned items. Second, three otherwise identical losses measure the effects of supervision and competition separately. Third, we give a practical rule for each generator: adopt appended-target training only when its adjusted development lower bound is positive. Experiments with released OneRec and local Amazon generators show that probability competition can be harmful. In the primary Amazon Video Games study, removing it improved all four point estimates by 7.8--22.2\%; held-out categories show that appended-target training helps some generators but should be rejected for others.

\section{Related Work}
\label{sec:related}

\paragraph{Generative recommendation and downstream reranking.}
Generative recommenders retrieve items by generating semantic IDs from a user's history~\cite{rajput2023tiger}; a recent survey reviews this design space~\cite{deldjoo2026generative}. Most work improves retrieval through semantic ID design, joint generation and ranking, sequence reranking, or reward-guided decoding~\cite{hua2023index,singh2024generalization,feng2026psg,sun2026unisgr,ren2024nonautoregressive,lin2025gref,song2026degre,yang2026rgd}. Other systems use a generator within a larger pipeline: it may replace an earlier cascade~\cite{wang2026unirec}, supply some candidates to downstream ranking~\cite{yang2026complementary}, or verify a frozen retriever's top-$K$ items~\cite{zhang2026verifier}. Related studies explain retrieval failures through item age, semantic ID quality, decoding difficulty, or early beam-search pruning~\cite{peng2026colditems,wang2026isd,yu2026daso,sun2026path}.

We study what happens after retrieval. Joint-prefix alignment changes generator supervision and inference-time completion~\cite{sun2026path}. We leave the generator and its returned candidates unchanged, then train the reranker with observed targets that the generator missed. Our question is whether those training-only targets improve or harm the ranking of candidates that can still be served.

\paragraph{Learning with a fixed candidate pool.}
Several areas study candidate distributions at other points in the pipeline. Cascade methods coordinate multiple ranking stages~\cite{gallagher2019joint,qin2022rankflow}. Selection-bias and off-policy methods use logged exposure to estimate relevance or retrieval-policy value~\cite{ovaisi2020selection,ovaisi2021propensity,wang2025offpolicy,betlei2026sidope}. Sampled-softmax methods correct the negative samples used to approximate catalog-level training~\cite{wu2024sampled}. Automated and calibration-aware rerankers choose strategies or balance relevance with preference calibration~\cite{gao2025llm4rerank,jeon2024calibration}. We keep the generator, serving candidates, and scorer architecture the same; only the use of missed observed targets during reranker training changes.

Machine-translation reranking provides a closely related finding: adding gold references to training candidates can create harmful comparisons~\cite{bhattacharyya2021energy}. We study the same train/test candidate mismatch in recommendation and separate the changes mixed by a simple append-versus-no-append comparison.

\paragraph{Separating the effects of training-only targets.}
ListNet provides the listwise probability model used by our losses~\cite{cao2007listnet}. We adapt profile minimization~\cite{murphy2000profile} to remove competition between returned and training-only items while preserving ranking within each group. Decoupled knowledge distillation (DKD) also separates learning within groups from competition between them, but its groups are target and non-target classes~\cite{zhao2022decoupled}. Our two groups differ in when they exist: returned items are ranked at inference, whereas appended items appear only during training. Multitask gradient methods~\cite{NEURIPS2020_3fe78a8a,chen2018gradnorm} balance task updates, while learning using privileged information (LUPI)~\cite{vapnik2009privileged,sharmanska2013ltrprivileged} uses training-only features. Neither compares otherwise identical losses with and without probability competition from training-only items. Table~\ref{tab:closest_work} summarizes these differences.

\begin{table}[t]
\caption{How the proposed comparison differs from nearby approaches. We keep the generator, inference pool, and scorer architecture fixed while changing one part of reranker supervision at a time.}
\label{tab:closest_work}
\centering\footnotesize
\setlength{\tabcolsep}{3pt}
\renewcommand{\arraystretch}{1.12}
\begin{tabular}{@{}>{\raggedright\arraybackslash}m{.37\columnwidth}>{\raggedright\arraybackslash}m{.55\columnwidth}@{}}\toprule
Approach & Difference from our study\\\midrule
Profile likelihood~\cite{murphy2000profile} & minimizes over a shared offset; does not compare these reranker losses\\\midrule
DKD~\cite{zhao2022decoupled} & splits target/non-target knowledge; we split returned and training-only items\\\midrule
Cascade optimization and calibrated reranking~\cite{gallagher2019joint,qin2022rankflow,ren2025scalr} & coordinates stages or calibrates sublists; our generator remains fixed\\\midrule
LUPI~\cite{vapnik2009privileged,sharmanska2013ltrprivileged} & uses training-only features, not training-only comparison items\\\midrule
Candidate-free verifier~\cite{zhang2026verifier} & learns token likelihood without completed-list supervision\\\midrule
Ours & compares three losses with the same generator, inference pool, and scorer architecture\\
\bottomrule\end{tabular}
\end{table}

For loss selection, we also include pairwise-ranking baselines within the returned pool or between appended targets and returned nontargets~\cite{rendle2009bpr}. Appendix~\ref{sec:soft_targets} extends the derivation to targets with different weights.

\section{Method}
\label{sec:design}

Our goal is to determine which change introduced by appended targets alters the ranking of the returned candidates. We first define the returned and appended item sets. We then split the completed-list loss into three terms and construct three training losses that add those terms one at a time.

\subsection{Returned and Missed Targets}

Let $x$ denote a user's interaction history and $Y_x$ the set of observed target items associated with it. The targets may be collected within a session or a fixed future window. For the ranking loss, membership in $Y_x$ is a binary target indicator; it records the observed event and does not necessarily indicate favorable sentiment.

A generative recommender retrieves an item by generating its semantic ID rather than scoring every catalog item. If item $i$ has semantic ID $t(i)=(t_1,\ldots,t_{L_i})$, the fixed generator assigns it log likelihood
\[
\ell(i\mid x)=\sum_{j=1}^{L_i}\log p_\theta(t_j\mid x,t_1,\ldots,t_{j-1}).
\]
The decoding temperature is fixed within each dataset. Let $\mathcal I$ be the items representable by the generator. Decoding returns the \emph{returned candidate set} $N(x)$. Some observed targets may be representable but absent from this set; we denote them by
\[
A(x)=(Y_x\cap\mathcal I)\setminus N(x).
\]
Candidate completion appends these missed targets during training, forming $C(x)=N(x)\cup A(x)$ without removing any returned item. At inference, the reranker still receives only $N(x)$; evaluation labels never alter this pool.

The reranker adds a learned residual to the fixed generator score:
\[
s_\phi(i\mid x)=\ell(i\mid x)+h_\phi(x,i).
\]
Here $h_\phi$ is a scalar correction computed from fixed user and item representations and candidate features. The attention variant also uses $N(x)$ as context. Its zero-initialized output layer recovers the generator ranking before training. We train only $\phi$ and resolve ties by complete item IDs.

For any candidate set $C$ containing at least one target, training uses uniform multi-target listwise cross-entropy:
\begin{equation}
\mathcal L(s,C,Y)=-\frac{1}{|Y\cap C|}
\sum_{i\in Y\cap C}\log\frac{e^{s_i}}{\sum_{j\in C}e^{s_j}},
\label{eq:loss}
\end{equation}
where $s_i=s_\phi(i\mid x)$ and $Y=Y_x$; we omit $x$ below. The target distribution has total mass one and divides it equally among the observed targets. Appendix~\ref{sec:soft_targets} gives the corresponding result for supplied nonnegative target weights.

\subsection{What Candidate Completion Changes}
\label{sec:loss_tasks}

Consider a training list with $r=|N\cap Y|>0$ retrieved targets and $k=|A|>0$ appended targets. Their shares of the target distribution are $\alpha=r/(r+k)$ and $\gamma=k/(r+k)$. Define the score normalizers $Z_N=\sum_{i\in N}e^{s_i}$ and $Z_A=\sum_{i\in A}e^{s_i}$, and normalize scores separately within each group: $p_N(i)=e^{s_i}/Z_N$ and $p_A(i)=e^{s_i}/Z_A$. This gives two within-group losses,
\[
\mathcal L_N=-\frac1r\sum_{i\in N\cap Y}\log p_N(i),\qquad
\mathcal L_A=-\frac1k\sum_{i\in A}\log p_A(i).
\]
The first ranks retrieved targets against the other returned items. The second encourages equal scores among the appended targets.

\paragraph{Proposition 1 (Three-part decomposition).}
The loss on the completed list separates into three terms:
\begin{equation}
\begin{split}
\mathcal L_{\mathrm{append}}&=\mathcal T_N+\mathcal T_A+\mathcal T_{\mathrm{mass}},\\
\mathcal T_N&=\alpha\mathcal L_N,\qquad
\mathcal T_A=\gamma\mathcal L_A,\\
\mathcal T_{\mathrm{mass}}&=-\alpha\log\frac{Z_N}{Z_N+Z_A}
-\gamma\log\frac{Z_A}{Z_N+Z_A}.
\end{split}
\label{eq:loss_tasks}
\end{equation}

\paragraph{Derivation.}
For a retrieved item, $e^{s_i}/(Z_N+Z_A)=[Z_N/(Z_N+Z_A)]p_N(i)$; the same identity holds for an appended item. Splitting the target sum in Equation~\ref{eq:loss} gives the two conditional losses with weights $\alpha$ and $\gamma$, followed by the two group-probability terms in Equation~\ref{eq:loss_tasks}.

\paragraph{Interpretation.}
The completed-list loss $\mathcal L_{\mathrm{append}}$ is Full. Relative to unweighted training on $N$, Full makes three changes. First, $\mathcal T_N$ multiplies the returned-item loss by $\alpha$. Second, $\mathcal T_A$ trains the scorer to rank within the appended group. Third, $\mathcal T_{\mathrm{mass}}$ makes the returned and appended groups compete for total probability in the target ratio $(\alpha,\gamma)$. We refer to this third term as the group-competition term. When $k=1$, $\mathcal T_A$ is zero because the appended group contains only one item. In Figure~\ref{fig:training_pool}(a), $r=1$ and $k=2$, so $\mathcal T_{\mathrm{mass}}$ asks the appended group to receive $2/3$ of the target probability although it initially receives only $0.011\%$. We next remove this competition while retaining both within-group losses.

\begin{figure*}[t]
\centering
\includegraphics[width=\textwidth]{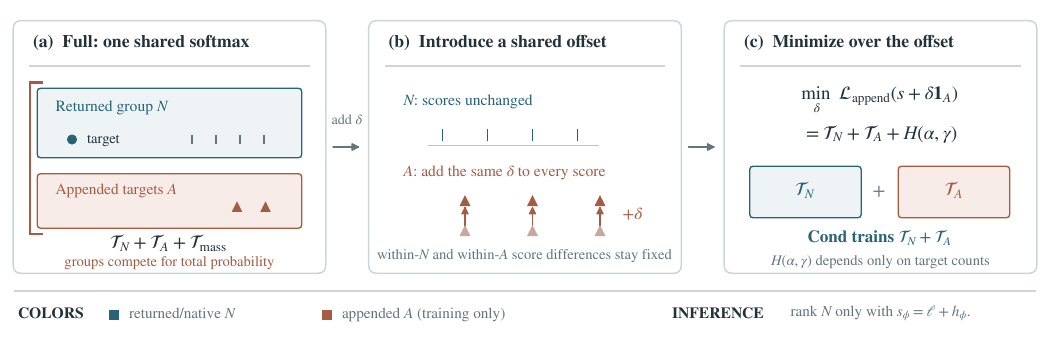}
\caption{Removing competition between the two groups. Blue denotes the returned group $N$ and orange the appended group $A$. (a) Full normalizes both groups together. (b) A temporary offset $\delta$ changes the appended group's total probability without changing rankings within either group. (c) Minimizing Full over $\delta$ leaves $\mathcal T_N+\mathcal T_A$ plus a count-dependent constant $H(\alpha,\gamma)$. This gives Cond. Inference ranks only $N$; the offset is not stored.}
\Description{Three stages show joint completed-list training, a common appended-score offset for each list, and analytical minimization over that offset. The resulting Cond loss retains the two within-group terms, and inference ranks only the returned candidates.}
\label{fig:audit_design}
\end{figure*}

\subsection{Conditional Training}
\label{sec:conditional_training}

We obtain Cond from Full by removing only the competition between the two groups. For each training list, temporarily add the same scalar $\delta$ to every appended score. This changes the total probability assigned to $A$ without changing any ranking within $N$ or $A$. The offset replaces $Z_A$ by $e^\delta Z_A$ while leaving $p_A$, $\mathcal L_A$, and all returned-item quantities unchanged. Differentiating Full with respect to $\delta$ gives
\[
\frac{e^\delta Z_A}{Z_N+e^\delta Z_A}-\gamma,
\]
which is zero at
\[
\delta^*=\log(k/r)+\log Z_N-\log Z_A.
\]
For the example in Figure~\ref{fig:training_pool}(a), $\delta^*\approx9.8$ moves the appended-group mass to its $2/3$ target while preserving both within-group rankings. Figure~\ref{fig:audit_design} illustrates this construction.

\paragraph{Proposition 2 (Removing group competition).}
For finite scores and $r,k>0$, the second derivative with respect to $\delta$ is the positive Bernoulli variance of the predicted appended-group mass. Hence $\delta^*$ is the minimum and
\[
\min_\delta\mathcal L_{\mathrm{append}}(s+\delta\mathbf1_A)
=\mathcal T_N+\mathcal T_A+H(\alpha,\gamma),
\]
where $\mathbf1_A$ indicates the appended group and $H(\alpha,\gamma)=-\alpha\log\alpha-\gamma\log\gamma$. The entropy term depends only on target counts, so it has no gradient with respect to the scorer. We can therefore train Cond directly as $\mathcal T_N+\mathcal T_A$; the temporary offset is neither computed nor stored in the implementation. Inference still applies $s_\phi=\ell+h_\phi$ only to $N$.

Cond is unchanged if the same constant is added to all scores in either group. Its gradient is $\nabla_\phi(\mathcal T_N+\mathcal T_A)$ wherever the scores are differentiable: training updates the scorer through the two within-group losses, without a group-competition gradient. A fixed offset in Full cannot remove that gradient throughout training, because the optimal offset changes with the scores.

Two special cases determine which lists enter training. If $k=0$, the generator missed no target and the loss is $\mathcal L_N$. If $r=0$, the returned list contains no target whose ranking can be compared across the three losses; minimizing over $\delta$ approaches $\mathcal L_A$ as $\delta\to+\infty$. We therefore train the three losses on lists with $r,k>0$. Evaluation still includes every eligible user, including those with $r=0$. If $k=1$, $\mathcal T_A=0$ and Cond reduces to $\mathcal T_N$. Appendix~\ref{sec:multisource} extends the construction to several candidate sources by using one temporary offset per source.

\subsection{Three Training Losses}
\label{sec:matched_objectives}

We train a separate reranker with each of three losses. Moving from one loss to the next adds exactly one requirement:
\begingroup
\setlength{\abovedisplayskip}{3pt}
\setlength{\abovedisplayshortskip}{3pt}
\setlength{\belowdisplayskip}{3pt}
\setlength{\belowdisplayshortskip}{3pt}
\[
\begin{aligned}
\mathcal L_{\mathrm{WN}}&=\mathcal T_N=\alpha\mathcal L_N,
&\mathcal L_{\mathrm{Cond}}&=\mathcal T_N+\mathcal T_A,\\
\mathcal L_{\mathrm{Full}}&=\mathcal T_N+\mathcal T_A+\mathcal T_{\mathrm{mass}}.
\end{aligned}
\]
\endgroup
WN uses only the returned-item loss, weighted exactly as it is in Full. Cond also learns to rank the appended targets, but keeps their probability separate from the returned group. Full then makes the two groups compete for probability. Candidates, training lists, scorer inputs, initialization, and optimization are unchanged. Table~\ref{tab:contrast_identification} shows the comparison for each question.
\label{sec:identification}

\begin{table}[h]
\caption{How the three trained rerankers answer separate questions. $V_W$, $V_C$, and $V_F$ are their ranking scores on the returned candidates; the first score is minus the second.}
\label{tab:contrast_identification}
\centering
\footnotesize
\setlength{\tabcolsep}{4pt}
\begin{tabular}{@{}lll@{}}
\toprule
Score difference & Only change & Question answered\\
\midrule
$V_C-V_W$ & add $\mathcal T_A$ & Does appended-target training help?\\
$V_C-V_F$ & remove $\mathcal T_{\rm mass}$ & Does competition hurt?\\
$V_F-V_W$ & add both terms & What is their combined effect?\\
\bottomrule
\end{tabular}
\end{table}

\paragraph{Why all three losses are needed.}
Let $V_W,V_C,V_F$ denote the returned-candidate ranking scores after training with WN, Cond, and Full. Comparing Full directly with WN gives
\[
V_F-V_W=(V_C-V_W)-(V_C-V_F).
\]
Full versus WN combines two changes, so a better Full score does not identify which change helped; one may help while the other hurts. Cond separates them. $V_C-V_W$ adds only the appended-target loss, while $V_C-V_F$ removes only group competition.

\subsection{Why the Effect Reaches Inference}

Although appended targets disappear at inference, group competition can still change the returned ranking. The same scorer produces both groups' scores during training, so $\mathcal T_{\mathrm{mass}}$ updates parameters that later rank returned candidates. Let $\pi_A=Z_A/(Z_N+Z_A)$ and $\mu_B=\sum_{i\in B}p_B(i)\nabla_\phi s_i$ for $B\in\{N,A\}$. Since $\nabla_\phi\log Z_B=\mu_B$,
\begin{equation}
\nabla_\phi\mathcal T_{\mathrm{mass}}
=(\pi_A-\gamma)(\mu_A-\mu_N).
\label{eq:mass_gradient}
\end{equation}
In Equation~\ref{eq:mass_gradient}, the first factor is the gap between the appended group's predicted probability and its target share. The second is the difference between the two groups' average score gradients. Their product updates the same scorer parameters later used to rank returned candidates. Group competition can therefore change inference rankings even though no appended item is served.

\section{Experiments}
\label{sec:experiment_design}

\subsection{Research Questions}

We ask three research questions (RQs). \textbf{RQ1:} How does probability competition between returned and appended items affect the ranking of returned candidates? \textbf{RQ2:} When that competition is removed, does supervision from appended targets improve the returned ranking? \textbf{RQ3:} Can development data choose between returned-only and appended-target training for a new category?

\begin{figure*}[t]
\centering
\includegraphics[width=\textwidth]{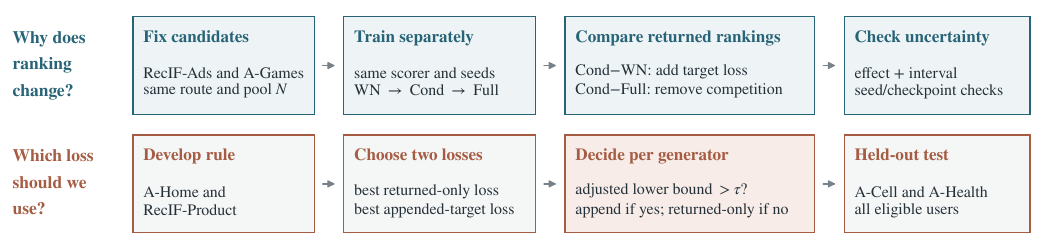}
\caption{Experiments for the three research questions. Top: RQ1 compares Cond with Full, and RQ2 compares Cond with WN, after separately training all three losses with identical settings. Bottom: RQ3 selects the best returned-only and appended-target losses on development users, switches only when the adjusted lower bound of the gain is above zero, and evaluates the chosen loss on held-out A-Cell and A-Health users.}
\Description{The upper lane separately trains WN, Cond, and Full with the same candidate route, scorer, and seeds, then reports the differences in their ranking metrics. The lower lane selects the best returned-only and appended-target losses on development users, switches only when the adjusted lower bound of the gain is positive, and evaluates the selected loss on held-out users.}
\label{fig:decision_gate}
\end{figure*}

Figure~\ref{fig:decision_gate} connects each question to a test. RecIF-Ads, using a released OneRec checkpoint, separates the effect of appended-target supervision from that of group competition. A comparison fixed before the Amazon Video Games (A-Games) final labels were opened provides a second test of competition. We also shift every appended score by the same amount. This preserves rankings inside each group but changes how probability is divided between them; a response from Full but not Cond directly identifies competition as the cause. Amazon Digital Music (A-Music) and Toys and Games (A-Toys) test whether the conclusions depend on one generator or reranker.

The remaining experiments address the training decision rather than another isolated loss term. Amazon Home and Kitchen (A-Home) and the RecIF-Bench Product split (RecIF-Product) establish a conservative rule: for each generator, switch from the best returned-only loss to the best appended-target loss only when the adjusted lower end of the development gain is positive. We fix this rule before training on Amazon Cell Phones and Accessories (A-Cell) and Amazon Health and Personal Care (A-Health). New generators and previously unseen users in those categories then show whether the rule chooses well. Table~\ref{tab:fitting_population} states the role of every dataset.

\subsection{Data and Candidate Generators}

Across all datasets, observed targets missed by the fixed generator are appended only during training; evaluation ranks the generator's original candidates. Comparing WN, Cond, and Full requires a training list with at least one retrieved and one missed target ($r,k>0$), and the appended-target loss is nonzero only when at least two targets were missed ($k\geq2$). All losses in a comparison use the same eligible lists. Evaluation retains every user, including those for whom the generator retrieved no target ($r=0$). Table~\ref{tab:fitting_population} reports how often a target is reachable. Throughout the experiments, \emph{returned-only} means any loss trained without appended targets; the capitalized \emph{Native} denotes the specific unweighted $\mathcal L_N$ baseline.

The two local generator architectures are Transformers and gated recurrent units (GRUs). A generator label includes its architecture and training seed; for example, ``Transformer (seed 42)'' denotes one trained generator and its resulting candidate pools.

\begin{table*}[t]
\caption{Datasets, candidate generators, and the question each addresses. The second column counts training lists containing both a returned and a missed target. The third shows how often at least two targets are appended, so their within-group loss is nonzero. The fourth shows how often the final returned candidates contain any target; ranges span generators.}
\label{tab:fitting_population}
\centering
\footnotesize
\setlength{\tabcolsep}{3pt}
\begin{tabularx}{\textwidth}{@{}>{\raggedright\arraybackslash}Xrrr>{\raggedright\arraybackslash}X@{}}
\toprule
Dataset / candidate generator & \shortstack{Training lists with both\\returned and missed targets} & \shortstack{At least two\\appended targets} & \shortstack{Final users with a\\returned target} & Question addressed \\
\midrule
\multicolumn{5}{@{}l}{\textit{Tests of the main claims}}\\
RecIF-Ads / OneRec-1.7B-Pro & 3,189 & 81.6\% & 44.9\% & separate supervision from competition \\
A-Games / Transformer (seed 42) & 914 & 75.5\% & 36.3\% & test competition on held-out labels \\
A-Home / GRUs (seeds 91--93) & 461--543 & 91.9--93.6\% & 9.6--10.3\% & determine when to switch losses \\
RecIF-Product / OneRec-1.7B & 2,224 & 100.0\% & 30.3\% & check whether training improves the initial ranking \\
A-Cell / GRUs (seeds 111--113) & 394--400 & 94.7--95.9\% & 30.1--30.7\% & evaluate the fixed switching rule \\
A-Health / GRUs (seeds 121--123) & 666--702 & 96.7--97.0\% & 19.9--20.8\% & evaluate the fixed switching rule \\
\midrule
\multicolumn{5}{@{}l}{\textit{Tests across other models and data}}\\
A-Music / Transformer (seed 42) & 524 & 63.5\% & 58.1\% & test a second data source \\
A-Games / Transformer (seed 42), GRUs (seeds 51--52) & 914--956 & 74.0--75.9\% & 36.3--38.6\% & vary the generator and reranker \\
A-Toys / GRUs (seeds 61--62) & 163--170 & 86.7--92.2\% & 13.6\% & test another category \\
\bottomrule
\end{tabularx}
\end{table*}

The six Amazon studies use public review histories~\cite{mcauley2015styles}. After retaining a user's first review of each item, the last three distinct items form the future target set and at least two earlier items form the history. This construction yields many users with both retrieved and missed targets. A robustness check keeps only targets rated at least four, distinguishing favorable feedback from interaction alone.

RecIF-Bench temporally separates each user's history from a released set of future clicks~\cite{zhou2026openonerec}. We use its Ads and Product splits. Each row defines one training group, and its deduplicated future clicks form $Y_x$. RecIF-Ads contains 3,740 reranker-training groups with a reachable target; the 3,189 groups with $r,k>0$ enter the three-loss fits. Its groups contain 1--10 targets (median 3, mean 3.75), with $r=1$--7 and $k=0$--9. RecIF-Product requires $r>0,k\geq2$ for fitting; its 2,224 groups contain 3--10 targets (median 8), with $r=1$--5 and $k=2$--9. Figure~\ref{fig:training_pool}(a) illustrates candidate completion for one RecIF-Ads training group.

Hash-based user splits prevent the same user from entering generator training, reranker training, development, and final evaluation. For A-Music and A-Games, we fixed data construction and training before obtaining the data. A-Games also removes 485 A-Music users, and its predictions and checkpoint choices were complete before final targets were opened. A-Home excludes users from the earlier Amazon categories, while RecIF-Product keeps final users out of training and development. Appendix~\ref{sec:gru_protocol} gives the exact split rules.

\paragraph{Generator models.}
\label{sec:generators}
The two RecIF splits use released OneRec checkpoints: OneRec-1.7B-Pro for RecIF-Ads and OneRec-1.7B for RecIF-Product. For RecIF-Ads, we project the fixed states to 128 dimensions and decode at temperature 1.2. The local Transformers and GRUs use training-only collaborative embeddings, residual quantization, beam width 64, and temperature 1. Validation negative log-likelihood (NLL) selects a checkpoint. Each generator seed rebuilds candidate pools, representations, and eligible training groups, so each dataset--generator pair is analyzed separately. Appendix~\ref{sec:gru_protocol} gives architectures and training settings.

\subsection{Isolating Supervision and Competition}
\label{sec:scorers}

To answer RQ1--RQ2, we train separate WN, Cond, and Full rerankers and evaluate them on the same returned candidates. They share the candidates, training lists, features, initialization, minibatch order, and optimizer. Cond minus WN therefore measures what changes when the appended-target loss is added. Cond minus Full measures what changes when group competition is removed but both within-group losses remain. We also compare WN with WN$+$Mass, which adds competition to WN without adding the appended-target loss. If both comparisons favor removing competition, the result does not depend on whether the appended-target loss is present.

We test both a multilayer perceptron (MLP) with hidden widths 256 and 128 and a two-block attention reranker with width 128 and four attention heads. A \emph{training run} means one independently initialized and trained reranker; it is not an attention head. Both rerankers use the same unchanged query and item representations and scalar features. The attention model attends only to returned candidates, so adding training-only targets cannot change its input context. Appendix~\ref{sec:metrics_scorer} specifies the inputs and architectures.

All rerankers use AdamW, batch size 64, weight decay $10^{-4}$, and gradient clipping at norm 1. The primary A-Games test uses three independent runs, 60 epochs, learning rate $10^{-4}$, and checkpoints chosen on development users. RecIF-Ads first uses three runs and then seven new runs for WN, Cond, Full, and a Full variant initialized with the correct appended-group probability. Training for 240 epochs and selecting checkpoints in five folds test whether the conclusion depends on stopping at epoch 60. Appendix~\ref{sec:equal_tuning_protocol} gives the schedules and defines the average effect over training.

\paragraph{Other generators and rerankers.}
Six further comparisons combine MLP and attention rerankers, Transformer and GRU generators, and the A-Games and A-Toys datasets. Within each comparison, every loss receives the same hyperparameter search; the mean development score over three training runs chooses the schedule for continued training. Appendix~\ref{sec:equal_tuning_protocol} lists the seeds, search ranges, and stopping rules.

\paragraph{Testing probability competition.}
\label{sec:mechanism_design}
RecIF-Ads and two Amazon comparisons test whether Full responds to the probability divided between returned and appended items. For each training list, we add the same constant to every appended score so that the appended group starts with its target probability $\gamma$; the Amazon comparisons also test half this shift. This operation does not change rankings inside either group and leaves Cond unchanged. Candidates, features, attention context, training budget, and inference are identical, so a change in Full comes from probability competition during training.

\subsection{Choosing Whether to Append Targets}

RQ3 asks whether development results can determine whether appended-target training should be used for a new category. For each generator, we first select the best returned-only loss: Native and WN, or an expanded set that also includes the native pairwise loss (NP). We separately select among Cond, Full, and a discriminative appended-target loss (Disc). Disc trains appended targets to outrank returned nontargets without putting the two groups in one softmax. Appendix~\ref{sec:disc_detail} defines Disc and NP formally.

For each generator, we train every loss for the same budget. Mean development full-target normalized discounted cumulative gain at 20 (FT-NDCG@20) chooses one returned-only loss and one appended-target loss. A-Home and RecIF-Product are the only datasets used to set the rule. Before training on A-Cell and A-Health, we specify the data construction, available losses, and switching threshold. Eight training runs choose the two losses on development users. We adopt the appended-target loss only if the lower confidence bound of its gain remains above zero after correcting for three generator comparisons; otherwise we keep the returned-only loss. Five new runs evaluate that choice on 5,398 A-Cell and 13,622 A-Health final users. Each category uses three newly trained GRU generators. Adjusting all 18 possible loss-pair comparisons leaves every choice unchanged (Appendix~\ref{sec:gru_protocol}).

\subsection{Metrics and Uncertainty}
\label{sec:prospective_uncertainty}

FT-NDCG@20 is our primary end-to-end metric. The reranker orders only returned candidates, but the ideal-ranking denominator contains the user's complete eligible target set. A missed target therefore remains in the denominator, and a user with no reachable target receives zero. Candidate-conditional NDCG instead removes unreachable targets from its denominator, so its values are not numerically comparable with FT-NDCG@20. Recall@20 and mean reciprocal rank at 20 (MRR@20) provide secondary checks. Appendix~\ref{sec:metrics_scorer} also reports results restricted to users with a reachable target.

We set the FT-NDCG improvement threshold to $\tau=0$ before held-out evaluation. The lower bound uses variation across reranker training runs on the same development users; separate final users evaluate the chosen loss. An application may instead require a positive minimum useful gain. Appendix~\ref{sec:gru_protocol} reports results for several thresholds. If the untrained initialization scores best on development users, as on RecIF-Product, we retain it.

For A-Games, one interval resamples the 9,121 final users while holding the three trained rerankers fixed; a second interval measures variation across those training runs.

RecIF-Ads intervals resample its evaluation users. For A-Cell and A-Health, we first average each user's predictions over three generators and five new training runs. The main intervals resample users; an additional analysis also resamples generators and training runs. Every loss comparison includes all final users. Appendix~\ref{sec:gru_protocol} gives the exact splits, seeds, search ranges, jointly tested comparisons, and resampling algorithms.

\section{Results}
\label{sec:confirmation}

The results answer the questions in causal order. We first isolate the harm from group competition, then measure the benefit of appended-target supervision, and finally test whether development results avoid harmful switches on new categories.

\begin{table*}[t]
\caption{RecIF-Ads FT-NDCG changes ($10^{-3}$). Each row is the first loss minus the second; positive values favor the first. Columns report the fixed epoch-60 result, the average effect across epochs 0--60, and the result after development folds select the checkpoint. Brackets are 95\% intervals over runs and users.}
\label{tab:onerec_inference}
\centering
\small
\setlength{\tabcolsep}{2.5pt}
\begin{tabular}{@{}lcccc@{}}\toprule
Training-loss change & \shortstack{Initial 3 runs\\epoch 60} & \shortstack{New 7 runs\\epoch 60} & \shortstack{New 7 runs\\average, epochs 0--60} & \shortstack{New 7 runs\\development-selected checkpoint}\\\midrule
All completion changes (Full $-$ WN) & $-6.97$ $[-10.22,-3.67]$ & $-5.55$ $[-8.07,-2.99]$ & $-11.80$ $[-13.73,-9.92]$ & $-1.92$ $[-4.02,+0.97]$ \\
Remove group competition (Cond $-$ Full) & $+8.63$ $[+6.20,+11.27]$ & $+7.42$ $[+4.98,+9.74]$ & $+12.41$ $[+10.31,+14.51]$ & $+1.03$ $[-1.24,+3.36]$ \\
Add appended-target loss (Cond $-$ WN) & $+1.66$ $[-0.50,+4.07]$ & $+1.88$ $[+0.34,+3.45]$ & $+0.60$ $[-0.20,+1.44]$ & $-0.89$ $[-2.21,+1.21]$ \\
\bottomrule\end{tabular}
\end{table*}

\begin{table*}[t]
\caption{Choosing whether to use appended-target training ($10^{-3}$ FT-NDCG). Each question states the subtraction; a positive change favors its first option. We switch from returned-only training only when the adjusted lower bound of the development gain is positive.}
\label{tab:update_decisions}
\centering\footnotesize
\setlength{\tabcolsep}{3pt}
\begin{tabularx}{\textwidth}{@{}lXrX@{}}\toprule
Dataset & Question and comparison & Change [95\% interval] & Conclusion\\\midrule
\multicolumn{4}{@{}l}{\emph{Choosing the switching rule}}\\
RecIF-Ads & After checkpoint selection, does Cond beat WN? & $-0.89$ [$-2.21,+1.21$] & No; keep WN\\
A-Games & On new users and runs, does Full beat WN? & $+1.75$ [$+0.48,+3.02$] & Yes; retain appended-target training as an option\\
A-Home & Does chosen appended training beat returned-only training? & $+0.21$ [$+0.12,+0.29$] & Yes; switch only when the adjusted lower bound is positive\\
RecIF-Product & Does the chosen trained loss beat the untrained scorer? & $0.00$ [$0,0$] & No; keep the untrained scorer\\\midrule
\multicolumn{4}{@{}l}{\emph{Applying the rule to new categories}}\\
A-Cell & Does the rule's choice beat returned-only training? & $+0.32$ [$-0.08,+0.72$] & It switched 2 of 3 generators; the overall gain is uncertain\\
A-Health & Does the rule's choice beat returned-only training? & $0.00$ [$0,0$] & It kept returned-only training for all 3 generators\\
A-Health & Does always appending beat returned-only training? & $-0.64$ [$-0.88,-0.41$] & No; the rule avoided a 1.7\% loss\\
A-Health & Does the lower-bound rule beat choosing the highest mean? & $+0.04$ [$-0.08,+0.17$] & No clear difference\\
\bottomrule\end{tabularx}
\end{table*}

\subsection{Group Competition Can Hurt}
\label{sec:prospective_results}
With the same 60-epoch budget, Cond ranked returned RecIF-Ads candidates better than Full in all seven new training runs. WN, Cond, and Full averaged $0.031527$, $0.033405$, and $0.025982$ FT-NDCG. Cond exceeded Full by $0.007423$ $[0.004978,0.009744]$ at epoch 60, and its average advantage throughout training was $+0.01241$ $[+0.01031,+0.01451]$. These intervals capture variation across training runs for the same evaluation users.

The difference depends on when training stops. When development folds chose the checkpoint, the interval crossed zero (Table~\ref{tab:onerec_inference}). After 240 epochs, Cond$-$Full was $-0.00064$ and $+0.00270$ at learning rates $3\times10^{-4}$ and $10^{-4}$; choosing the best among 25 recorded checkpoints gave $+0.00120$ and $+0.00108$. The advantage is consistent at epoch 60 but is not guaranteed under a different training duration or checkpoint rule.

Giving Full the correct appended-group probability at initialization raised its FT-NDCG by $0.009233$ $[0.006838,0.011643]$, with the same direction in all seven runs. Cond and WN were unchanged. This direct response to group probability supports probability competition as the source of Full's change.

\begin{table}[t]
\caption{A-Games FT-NDCG gain from removing competition ($10^{-3}$). Cond minus Full retains the appended-target loss; WN minus WN$+$Mass omits it. The 95\% intervals resample users and runs separately.}
\label{tab:games_primary}
\centering\footnotesize
\setlength{\tabcolsep}{1.8pt}
\begin{tabularx}{\columnwidth}{@{}lXrrr@{}}\toprule
Reranker & Loss comparison & Gain & Users only & Runs only\\\midrule
MLP & Cond minus Full & 9.177 & [6.709,11.651] & [8.202,10.152]\\
MLP & WN minus WN$+$Mass & 5.834 & [4.284,7.386] & [3.462,8.206]\\
Attention & Cond minus Full & 3.234 & [1.848,4.651] & [1.275,5.192]\\
Attention & WN minus WN$+$Mass & 5.470 & [3.886,7.111] & [$-0.824$,11.764]\\
\bottomrule\end{tabularx}
\end{table}

\paragraph{Confirmation on Amazon Video Games.}
We chose the A-Games generator, comparisons, learning rate, and training runs before opening the final labels. Removing competition improved FT-NDCG for both rerankers, with or without the appended-target loss, over 9,121 separate users (Table~\ref{tab:games_primary}). Relative gains were $22.23\%$ and $12.84\%$ for the MLP and $7.83\%$ and $13.16\%$ for attention. All four user intervals and three of four training-run intervals excluded zero, and every run favored removing competition.

\paragraph{Direct test of probability competition.}
\label{sec:mechanism_results}
The derivation predicts that shifting every appended score by the same amount leaves Cond unchanged. Across 12 runs, Cond's training path and learned parameters were identical to numerical precision. Full changed: for the A-Games attention reranker, the shift moved Cond$-$Full from $-0.000358$ to $+0.004049$, a change of $+0.004407$ $[0.001591,0.007177]$. Full therefore responds to the probability divided between the two groups, whereas Cond does not.

\subsection{Appended-Target Supervision Can Help}
\label{sec:loss_factorial}
Cond differs from WN only by the separately normalized loss over appended targets. After 60 epochs, Cond$-$WN was $+0.001877$ $[0.000343,0.003455]$ across the seven new RecIF-Ads runs, and every run had the same sign. When development folds chose the checkpoint, the estimate was $\OneRecCVCondWNPoint$ $[\OneRecCVCondWNLo,\OneRecCVCondWNHi]$ (Table~\ref{tab:onerec_inference}). Appended-target supervision helped at epoch 60, but its benefit was uncertain after checkpoint selection.

\subsection{Development Results Prevent a Harmful Switch}
\label{sec:home_confirmation}

Table~\ref{tab:update_decisions} summarizes the choices. On A-Cell, development users chose Disc for two generators and WN for the third. The resulting FT-NDCG change was $+.000319$ $[-.000077,+.000720]$; an interval that also resamples generators and training runs was $[-.000431,+.001309]$. On A-Health, development users kept a returned-only loss for every generator. Always forcing the best appended-target loss would have reduced FT-NDCG by $0.000644$ $[0.000412,0.000881]$, or 1.7\%. The rule therefore allowed appended-target training on A-Cell and avoided a clear loss on A-Health, although the A-Cell gain remained uncertain.
\label{sec:product_decision}

\section{Discussion}
\label{sec:discussion}

Appended targets affect inference through the scorer parameters shared with returned candidates. Full also trains the scorer to divide probability between returned and appended items, although only returned items appear at inference. We tested this competition by adding the same constant to every appended score in each training list. This changed Full's returned-item rankings but left Cond unchanged, showing that probability competition during training can alter the rankings used at inference.

Separating the two effects changes how candidate completion should be interpreted. Removing competition consistently helped in the primary A-Games study and after the same 60-epoch RecIF-Ads budget. The appended-target ranking task also helped after 60 epochs, but its estimated benefit weakened when development data chose the checkpoint. Thus, a poor result from appending targets does not imply that the extra labels are useless: their supervision may help while probability competition harms the same scorer. Comparisons must also use the same training budget. A-Home further shows that the conclusion depends on which interactions are treated as targets.

This distinction also leads to a practical recommendation. For each generator, train every candidate loss on the same candidate pool. On development users, select the best reranker trained without appended targets and the best reranker trained with them, then compare the two. A favorable isolated effect is insufficient because the deployed loss combines several effects. A-Cell and A-Health required different choices, showing why candidate completion should be decided for each generator rather than treated as a universal default.

\section{Conclusion}
\label{sec:conclusion}

Appending a missed target is not a neutral way to add supervision. It changes the weight on retrieved targets, adds a new ranking task, and makes training-only targets compete with candidates that will actually be ranked at inference. Our three-loss comparison separates these changes under the same scorer. It shows that group competition, rather than the extra supervision itself, can be the main source of degradation: removing competition improved all four prespecified A-Games comparisons by 7.8--22.2\%. Appended-target supervision helped RecIF-Ads after 60 epochs, although the estimated benefit became uncertain when development data selected the checkpoint.

For each generator, train returned-only and appended-target rerankers for the same inference pool and compare them on development users. Use appended-target training only when the adjusted lower confidence bound exceeds a chosen threshold. On held-out categories, this rule avoided a 1.7\% A-Health loss while allowing appended-target training for two A-Cell generators. Candidate completion should therefore be tested for each generator rather than applied automatically.
\label{sec:maintext_end}

\begin{acks}
This work was supported by the National Natural Science Foundation of China (Grant No. 62672438).
\end{acks}

\appendix
\section{Reproducibility and Additional Results}
\label{sec:implementation}
\setlength{\abovedisplayskip}{4pt}
\setlength{\abovedisplayshortskip}{3pt}
\setlength{\belowdisplayskip}{4pt}
\setlength{\belowdisplayshortskip}{3pt}

This appendix provides the details needed to reproduce the data splits, losses, model choices, and final evaluation. Table~\ref{tab:home_objectives} checks whether the conclusions change when targets are restricted to favorably rated items or when uncertainty includes both users and training runs.

\paragraph{Code, data, and ethics.}
\ifarxivversion
The arXiv ancillary archive contains the code, environments, split manifests, and commands needed to reproduce our results, with links to the public datasets and OneRec checkpoints.
\fi
We use only public secondary data, exclude review text and demographic attributes, and release no raw user data.

\subsection{Metrics and Scorer Implementation}
\label{sec:metrics_scorer}
Let $\pi_x(j)$ be the item at rank $j$. We compute
\begin{equation}
\operatorname{FT\text{-}NDCG@20}(x)=
\frac{\sum_{j=1}^{20}\mathbf1\{\pi_x(j)\in Y_x\}/\log_2(j+1)}
{\sum_{j=1}^{\min(20,|Y_x|)}1/\log_2(j+1)}.
\label{eq:ndcg}
\end{equation}
Equation~\ref{eq:ndcg} uses the complete eligible target set in its denominator. The candidate oracle ranks available targets first. Users with no reachable target remain in every all-user mean, and Table~\ref{tab:fitting_population} reports the percentage with a reachable target.

Recall@20 is the fraction of $Y_x$ in the top 20; MRR@20 is the reciprocal rank of its first item, or zero if absent. Both retain targets outside the returned pool. After 60 epochs on the seven new RecIF-Ads seeds, Cond$-$WN changed Recall/MRR by $+0.00166/+0.00425$, Cond$-$Full by $+0.00880/+0.01348$, and Full$-$WN by $-0.00713/-0.00923$, matching FT-NDCG.

The scorer root-mean-square normalizes query and item states with a $10^{-6}$ floor, then concatenates them with their product, absolute difference, and seven scalars into 519 dimensions. Slots 1/3 hold standardized log likelihood/training frequency; slots 2/4--6 are reserved zeros and slot 7 marks validity. Source membership is not a feature. Attention uses returned-candidate keys and values; output layers start at zero.

\subsection{Multiple Candidate Sources}
\label{sec:multisource}
Let disjoint groups $S_1,\ldots, S_G$ contain positive target counts $m_g$, let $\gamma_g=m_g/\sum_hm_h$, and define $Z_g=\sum_{i\in S_g}e^{s_i}$, $P_g=Z_g/\sum_hZ_h$, and the within-source loss $\mathcal L_g$. The same substitution used in Proposition~1 gives
\[
\mathcal L=\sum_{g=1}^G\gamma_g\mathcal L_g-\sum_{g=1}^G\gamma_g\log P_g.
\]
Add an offset $\delta_g$ to every score in $S_g$, fixing one because a common shift is redundant. The second term is cross-entropy from $\gamma$ to the source softmax, with gradient $P_g-\gamma_g$. Hence every finite minimum has $P_g=\gamma_g$ and value $H(\gamma)$; minimizing over all offsets leaves $\sum_g\gamma_g\mathcal L_g+H(\gamma)$. Cond is the $G=2$ case.

\paragraph{Soft targets.}
\label{sec:soft_targets}
For normalized nonnegative target weights $q_i$, let $\rho_B=\sum_{i\in B}q_i$ and $\mathcal L_B^q=-\sum_{i\in B}(q_i/\rho_B)\log(e^{s_i}/Z_B)$. With positive source masses,
\[
\begin{split}
\delta_q^*&=\log(\rho_A/\rho_N)+\log Z_N-\log Z_A,\\
\min_\delta\mathcal L_q(s+\delta\mathbf1_A)
&=\rho_N\mathcal L_N^q+\rho_A\mathcal L_A^q+H(\rho_N,\rho_A).
\end{split}
\]
Zero source mass uses the corresponding infinite-offset limit and omits the empty conditional term. Thus the same decomposition applies to any supplied nonnegative target distribution.

\begin{figure*}[t]
\centering
\includegraphics[width=\textwidth]{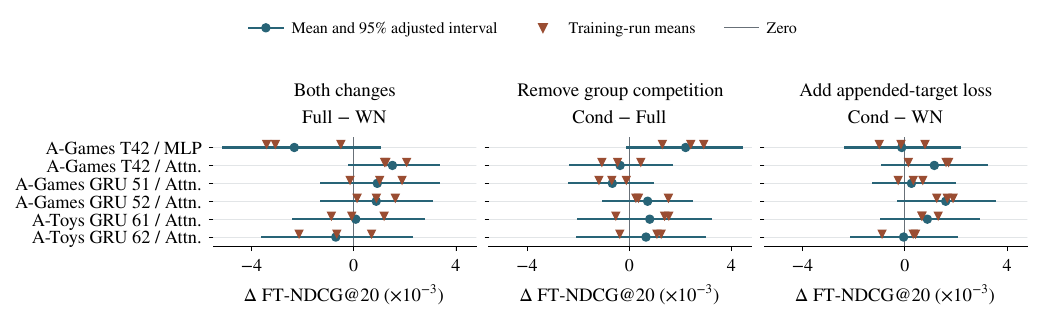}
\caption{Effect directions vary across six Amazon generator--reranker combinations, and most 95\% familywise-adjusted intervals include zero. T42 denotes the Transformer trained with seed 42; numeric labels denote GRU seeds. The legend identifies overall means and intervals, individual training-run means, and zero. All panels share their axes.}
\Description{Three forest-plot panels report Full minus WN, Cond minus Full, and Cond minus WN for six A-Games and A-Toys scorer-generator settings.}
\label{fig:local_attribution}
\end{figure*}

\begin{table*}[t]
\caption{FT-NDCG scores ($10^{-3}$). Each comparison is first loss minus second. ``Dev-selected'' means that development users choose the model; Reachable gives the number and percentage of final users with a returned target. The two score columns show first/second.}
\label{tab:support_effects}
\centering\footnotesize
\setlength{\tabcolsep}{1.5pt}
\begin{tabular*}{\textwidth}{@{\extracolsep{\fill}}lllrrrrr@{}}\toprule
Dataset & Comparison & Model & Reachable & All: scores & All: $\Delta$ & Reachable: scores & Reachable: $\Delta$ [95\% CI]\\\midrule
RecIF-Ads & Cond vs Full & Epoch 60 & 2504/5579 (44.9\%) & 33.40/25.98 & +7.42 & 74.43/57.89 & +16.54 [+11.38,+21.56] \\
A-Home & Cond vs Full & Dev-selected & 1943--2086/20323 (9.6--10.3\%) & 11.57/11.44 & +0.13 & 118.01/116.77 & +1.24 [-3.36,+6.85] \\
A-Cell & Selected vs returned-only & Dev-selected & 2327/5398 (43.1\% union) & 48.75/48.44 & +0.32 & 113.10/112.36 & +0.74 [-0.18,+1.66] \\
A-Health & Selected vs returned-only & Dev-selected & 3824/13622 (28.1\% union) & 38.04/38.04 & 0.00 & 135.49/135.49 & 0.00 [0,0] \\
\bottomrule\end{tabular*}
\end{table*}

\begin{table}[t]
\caption{A-Home results when targets are all interactions or only ratings of at least four ($10^{-3}$ FT-NDCG). Cond minus Full measures the effect of removing group competition; Cond minus WN measures appended-target supervision. ``Chosen appended minus returned-only'' compares the two losses selected on development users. The 95\% intervals resample users alone, then users, three generators, and five runs.}
\label{tab:home_objectives}
\centering
\scriptsize
\setlength{\tabcolsep}{1.5pt}
\begin{tabularx}{\columnwidth}{@{}>{\raggedright\arraybackslash}p{.18\columnwidth}Xrrr@{}}\toprule
Target definition & Loss comparison & Effect & Users only & Users + runs\\\midrule
\multirow{2}{.18\columnwidth}{All interactions} & Cond $-$ Full & $+0.128$ & $[-0.02,+0.28]$ & $[-0.34,+0.72]$ \\
& Chosen appended $-$ returned-only & $+0.072$ & $[-0.09,+0.23]$ & $[-0.58,+0.70]$ \\\midrule
\multirow{2}{.18\columnwidth}{Only ratings $\geq4$} & Cond $-$ Full & $-0.119$ & $[-0.27,+0.03]$ & $[-0.85,+0.77]$ \\
& Cond $-$ WN & $-0.232$ & $[-0.40,-0.07]$ & $[-0.83,+0.23]$ \\
\bottomrule\end{tabularx}
\end{table}

\subsection{Generators and Data Construction}
\label{sec:gru_protocol}
Local Transformers use 128-dimensional states, four encoder and decoder layers, four heads, feedforward width 1,024, and dropout .1; AdamW uses rate $10^{-4}$, decay .01, batch 512, and gradient norm 1. GRUs use two 128-dimensional encoder and decoder layers, dropout .1, and initial rate .001. Validation NLL selects epochs 12--80 (also 10--11 for A-Home); after eight epochs without .001 relative improvement, every two stale epochs halve the rate to a $1.25\times10^{-5}$ floor.

Training-prefix term-frequency--inverse-document-frequency (TF--IDF) embeddings reduced by singular value decomposition are 128-dimensional. Three 64-center residual-quantization levels plus a collision suffix define identifiers. Histories contain at most 20 items; beam-64 decoding forms $N$, and teacher forcing scores identifiers. Amazon tasks use the 2014 5-core archives~\cite{mcauley2015styles}; first reviews define targets, text is omitted, out-of-catalog events are removed before grouping, and final inputs omit targets.

The A-Games split excludes 485 A-Music users. For user $u$, $\operatorname{SHA\text{-}256}(\texttt{source-mass-v2-games:}+u)\bmod100$ assigns 0--19/20--34/35--44/45--99 to generator/fitting/development/final roles. The plan and predictions were fixed before final-target access. User intervals resample users while holding the three trained rerankers fixed; training-run intervals use $t_2=8.8602$. A further check resamples both.

All Amazon assignments use $\operatorname{SHA256}(s+u)\bmod100$. A-Music uses salt \texttt{source-mass-v1:} and ranges 0--49/50--69/70--79/80--99. A-Games, A-Toys, and A-Home use salts \texttt{source-mass-v2-games:}, \texttt{source-mass-v3-toys:}, and \texttt{source-mass-v4-confirmation:} with the A-Games ranges. Reranker selection uses seeds 42--44 and A-Home final evaluation 101--105. Figure~\ref{fig:local_attribution} summarizes the six settings.

Table~\ref{tab:support_effects} reports scores for all users and separately for users with a reachable target.

The 1,770,644,480-parameter OneRec-1.7B-Pro checkpoint represents a query by its last prompt state and an item by the mean of its three semantic ID token states. A fixed Gaussian matrix (seed 271828, $1/\sqrt{128}$ scaling) projects 2,048 dimensions to 128; full-prefix forward passes verify both state extraction and the temperature-1.2 path scores.

Let $h(u)$ map the first eight big-endian SHA-256 bytes to $[0,1)$. Salt \texttt{sidecar-openonerec-pro-\allowbreak ads-prospective-v1} and ranges $[0,.3)/[.3,.5)/[.5,1)$ assign reranker-training/evaluation/unused-test users (8,304/5,579 in the first two groups, with no overlap). Evaluation users form five hash-sorted round-robin folds. All 3,740 reachable training users (3,189 with $r,k>0$) and item-frequency features come only from the reranker-training group. Retrieved targets keep the maximum-likelihood path; additions use the smallest released product identifier. Re-inversion covers all 5,477 native and 11,453 added user--target pairs.

The first three runs established the comparison; seven new runs then used the same losses, epoch-60 endpoint, and evaluation users. Each of 5,000 draws from NumPy's PCG64 random-number generator (seed 20260916) resamples users within folds and the seven new runs. For every loss and run, the other four folds select the checkpoint and the held fold evaluates it before differences are averaged. The separate 240-epoch study uses all ten runs at rates $3\times10^{-4}$ and $10^{-4}$.

RecIF-Product uses the released OneRec-1.7B checkpoint and data revision \texttt{cc4d0b\allowbreak{}5b7294\allowbreak{}ecf75e\allowbreak{}40be1c\allowbreak{}77fa6b\allowbreak{}7d284b\allowbreak{}b84b}. Hash ranges $[0,.3)/[.3,.5)/[.5,1)$ under salt \nolinkurl{sidecar-openonerec-native-v1} assign 8,243 reranker-training, 5,668 development, and 13,999 final users. Seeds 42--44 screen rates $10^{-4},3\times10^{-4},10^{-3}$ for at most 60 epochs (patience 8); mean development FT-NDCG fixes loss-specific rates/checkpoints before final seeds 101--105, preferring lower rates and earlier epochs on ties. The 2,224 training groups satisfy $r>0,k\geq2$; all 9,758 final users with no reachable target remain. The three jointly tested loss comparisons use 20,000 PCG64-seed-20260926 draws and 98.3333\% percentiles.

The A-Home analysis holds out each of runs 101--105 and 301--303. The other seven choose among Cond, Full, and Disc, but replace the generator-specific returned-only loss only when a one-sided Student-$t$ lower bound, adjusted over three generators, is above zero. The paired-user interval averages the 24 decisions over 20,000 draws (PCG64 seed 20260918).

A-Cell uses salt \texttt{source-mass-v5-cell-confirmation}, generators 111--113, development runs 501--508, final runs 601--605, and 5,398 final users; users in earlier Amazon categories are excluded. A-Health uses salt \texttt{source-mass-v6-health-\allowbreak{}confirmation}, generators 121--123, development/final runs 701--708/801--805, and 13,622 final users. Unlike A-Cell, it does not filter user identifiers against earlier category experiments. Its four data roles remain disjoint, and no history, feature, or parameter transfers across categories. Both studies assign disjoint 20/15/10/55\% data roles, use a two-layer width-128 GRU with beam 64, and train attention rerankers for 480 epochs.

The fixed loss settings are WN cosine/.03, NP constant/.001, Cond/Full constant/.03, and Disc cosine/.1. A-Home selected NP for two generators and WN for one; it never selected the unweighted Native loss. The confirmation candidates therefore retain WN and NP rather than reopening the returned-only search. Mean development FT-NDCG over eight runs selects one returned-only loss and one appended-target loss. We switch to the latter only when the corrected lower confidence bound for their paired difference is positive; five new runs evaluate both losses. Rankings are fixed before final-target access. The development bound covers variation across runs for fixed development users. Primary final intervals use 20,000 paired-user draws; intervals that also resample generators and runs are reported as additional uncertainty checks.

For each category, we correct the three lower confidence bounds that compare the selected returned-only and appended-target losses, one comparison per generator. As a stricter check, we adjust all $3\times3\times2=18$ possible loss-pair comparisons ($t_7=3.9467$). The adjusted lower bounds for the selected pairs are $+.000411,-.001813,+.001278$ on A-Cell and $-.000713,-.000438,-.001317$ on A-Health, so the switch-or-retain decisions remain unchanged. Choosing solely by the highest mean development score gives the same decisions on A-Cell. On A-Health, that simpler approach selects Full only for generator 122 and yields $-0.000042$ $[-0.000167,+0.000082]$ versus the selected returned-only loss; retaining a loss unless its corrected lower bound is positive changes this result by $+0.000042$ $[-0.000082,+0.000167]$.

A post hoc check used margins $0$, $.00025$, $.0005$, and $.001$; these selected 4, 2, 1, and 1 switches to appended-target training. Their mean effects across settings are $+.000389$, $+.000273$, $+.000345$, and $+.000345$; their worst-setting effects are $-.000361$, $-.000361$, 0, and 0. Because this check was designed after outcomes were available, it shows how results vary with the threshold rather than confirming a new rule.

No evaluation interaction contributes to feature statistics.

\paragraph{Primary training configurations.}
RecIF-Ads uses a residual MLP, seeds 3101--3110, 60 epochs at $3\times10^{-4}$, and cross-validation over initialization and epochs 5,10,$\ldots$,60. A-Games uses MLP and attention rerankers, seeds 42--44, and 60 epochs at $10^{-4}$ including initialization. RecIF-Product uses a residual MLP, development/final seeds 42--44/101--105, three rates from $10^{-4}$ to $10^{-3}$, and patience eight.

\subsection{Optimization and Inference}
\label{sec:equal_tuning_protocol}
\begin{samepage}
The constant-rate grid is $\{3\!\times\!10^{-5},10^{-4},3\!\times\!10^{-4},10^{-3},3\!\times\!10^{-3},10^{-2}\}$; if the largest rate wins, the grid extends to $.03,.1,.3$. Cosine uses 24 warmup epochs, 480 total epochs, a 1\% floor, and the winning rate plus its neighbors. Ties prefer earlier checkpoints, then lower rates. Means over three development runs select among Native/WN, Native/WN/NP, or Cond/Full/Disc. For the user-and-run-averaged difference $d_j$ at $0=t_0<\cdots<t_m=T$, AULC is $T^{-1}\sum_{j=0}^{m-1}(t_{j+1}-t_j)(d_j+d_{j+1})/2$.
\end{samepage}

Full$-$WN and Cond$-$Full have different signs in two of six additional settings, showing that the combined comparison can hide the effect of group competition. Among ten intervals that resample both runs and users from five new runs, only A-Games Transformer (seed 42) Cond$-$WN excludes zero: $+0.001572$ $[+0.000137,+0.003305]$. Paired-user intervals also exclude zero for its Full$-$WN comparison and A-Toys GRU (seed 61) Cond$-$Full, when the trained rerankers are held fixed. Training losses were not reselected; Full's selected rate is interior in all six settings.

All model choices precede final evaluation. We adjust intervals jointly within three prespecified sets: 30 primary comparisons, 8 score-shift comparisons, and 10 comparisons from new training runs. Each uses 20,000 draws; intervals that include both users and training variation resample both users and the observed runs. OneRec repeats checkpoint choice inside \OneRecCVBootstrapDraws{} user-and-run draws; other local intervals use the saved predictions without choosing models again.

\subsection{Discriminative Control}
\label{sec:disc_detail}
Let $N^-=N\setminus Y$, with the pairwise term below defined as zero when $k|N^-|=0$. The discriminative control is
\begin{equation}
\mathcal L_{\mathrm{Disc}}=\mathcal T_N+\frac{\gamma}{k|N^-|}
\sum_{i\in A,j\in N^-}\log(1+e^{s_j-s_i}).
\label{eq:disc}
\end{equation}
Equation~\ref{eq:disc} ranks appended targets above returned nontargets with total weight $\gamma$, giving each target $\gamma/k=1/(r+k)$. At $r=0$ we set $\mathcal T_N=0$, so $k=1$ remains trainable when $|N^-|>0$. Inference still ranks $N$. The returned-item pairwise baseline applies the same loss to $i\in N\cap Y,j\in N^-$ and requires $r>0,|N^-|>0$.

\end{document}